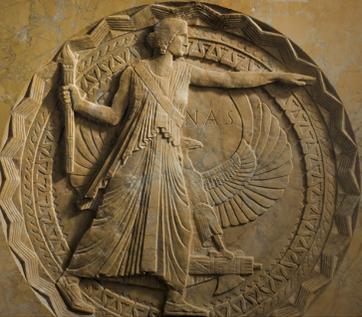

# A. G. W. Cameron
1925–2005

## BIOGRAPHICAL *Memoirs*

*A Biographical Memoir by*
*David Arnett*

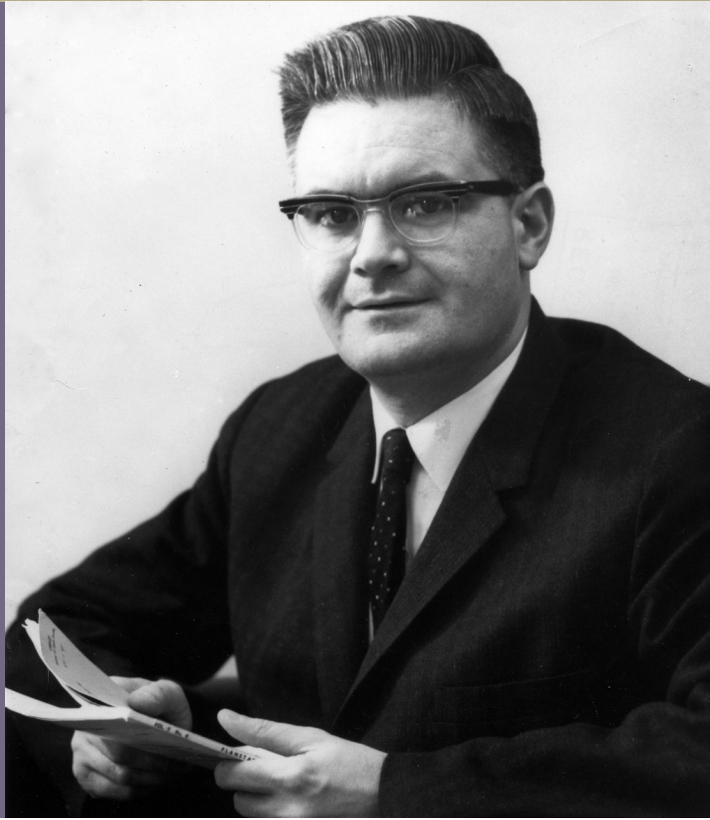

NATIONAL ACADEMY OF SCIENCES

# ALASTAIR GRAHAM WALKER **CAMERON**

*June 21, 1925–October 3, 2005*
Elected to the NAS, 1976

Alastair Graham Walker Cameron was an astrophysicist and planetary scientist of broad interests and exceptional originality. A founder of the field of nuclear astrophysics, he developed the theoretical understanding of the chemical elements' origins and made pioneering connections between the abundances of elements in meteorites to advance the theory that the Moon originated from a giant impact with the young Earth by an object at least the size of Mars. Cameron was an early and persistent exploiter of computer technology in the theoretical study of complex astronomical systems—including nuclear reactions in supernovae, the structure of neutron stars, and planetary collisions.

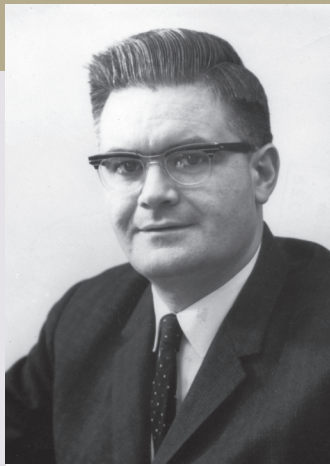

*By David Arnett*

Cameron was born in Winnipeg, Manitoba, to an academic family; his father was a professor of biochemistry at the Manitoba Medical College. "I was told that at the age of four I called all men 'Doctor,'" he recollected. "Clearly an early attempt at forming a hypothesis based upon limited data." Cameron studied mathematics and physics as an undergraduate at the University of Manitoba, worked summers at Canada's nuclear laboratory at Chalk River, and earned a Ph.D. in nuclear physics at the University of Saskatchewan. With Leon Katz as his advisor, his dissertation addressed photonuclear activation. At that time, nuclear physics was a hot topic, with a high rate of discovery: "By the time I defended my thesis, I had some 17 publications," Cameron later recalled (1).

Cameron was hired in 1952 as an assistant professor at Iowa State University (then College) at Ames to teach nuclear physics and help to support the new 70-MeV synchrotron. Tuning the new accelerator (to increase the beam current to an acceptable level for scientific research) was less than enthralling to Cameron, but one day in the library he found a new and very short paper—the observation by Paul Merrill of technetium in the spectra of red giant stars of class S—that was enthralling indeed.

Knowing that this element has only radioactive isotopes with likely half-lives of about 200,000 years—far less than the age of the star—Cameron reasoned that the star must have been producing technetium, most likely by a flood of neutrons.

This phenomenon was "clearly something very interesting and totally new," Cameron recalled, and he set out to investigate how it could happen. He began to teach himself astrophysics, buying all the graduate-level astrophysics texts he could find and learning how to calculate thermonuclear-reaction rates. One result was his first astrophysics paper (Cameron 1955), which identified the neutron source: the reaction in which an alpha particle ($He^4$ nucleus) was captured by a $C^{13}$ nucleus, emitting a neutron.

To the nonnuclear physicist, this discovery may have seemed like pulling a rabbit out of a hat—Cameron would be accused of this more than once; it seemed to be his style—but in fact the solution was more like solving a crossword puzzle. As Hans Bethe had shown (in his Nobel Prize-winning research), $C^{13}$ is a natural result of hydrogen-burning in the CNO cycle. After hydrogen has been burned, contraction and heating occurs, and helium-burning begins.

This process, which produces the nuclei $C^{12}$ and $O^{16}$, was unraveled by Edwin Salpeter and Fred Hoyle, who shared the Crafoord Prize for their work. Cameron introduced two distinct and controversial ideas: he identified the specific source of the neutrons, and the mixing of the neutron-exposed material on the stellar surface.

Cameron realized that the very large (stellar astronomy) and the very small (nuclear physics) were intimately related, but also that he had a lot more to learn—especially in the latter arena. "I had pretty much absorbed what I needed to know on the astrophysical side, but nuclear physics was still in a state of rapid development, so I applied to go back to Chalk River," which could provide the broader nuclear physics environment he needed. It would also give him access to computer resources, primitive as they were by today's standards.

Cameron knew that heavier elements had larger cross-sections for neutrons, so that neutrons would be preferentially captured by them, and he guessed that the relatively abundant element iron (26 protons) would be the "seed nucleus" to make still heavier nuclei, including technetium. The idea was straightforward, but the computation was too complex to be solved by traditional mathematical techniques; he would need to use computers to solve these and other complex networks of nuclear reactions.

As Cameron later described in (1,3,4), he began by programming an IBM accounting machine, which used Hollerith punched cards, to solve the helium-burning reaction network, by which $He^4$ is converted to $C^{12}$, $O^{16}$, and $Ne^{20}$. He graduated to an IBM 650, which used cards and magnetic drum (an accounting machine in Ottawa that would be on exhibit for a few weeks), and proceeded to solve a slow neutron capture network (now called the "s-process"). This gave quantitative predictions for the "cosmic abundances" of Suess and Urey (1956), which were based upon isotopic ratios measured in meteorites.

The computer at Chalk River was upgraded to a Datatron 205 with decimal coding, which was used during the day for accounting. To get access to the machine, Cameron "went on the night shift for three years," a story familiar to users of many hours of computer time. This led to a burst of publications, around 1959, involving reaction networks for carbon fusion, oxygen fusion, photo-disintegrations, and neutron stars, and also to Cameron's Chalk River report (Cameron 1957a). Not all this material appeared in refereed publications; a notable example was the "approach to nuclear statistical equilibrium" (now called silicon burning), which was described in the Yale lecture notes (5).

In 1959, Cameron felt that his dip into nuclear physics at Chalk River had run its course, and he accepted an invitation from Jesse Greenstein to spend a year in the astronomy department at Caltech, where he could immerse himself in astrophysics. Cameron was an unusual visitor for the time; he had worked on neutron stars (which pleased Fritz Zwicky) and did not ask to use the Palomar 200-inch telescope. He had not been at Caltech long when John Reynolds announced the discovery of the extinct radioactivity $I^{129}$, via its decay product $Xe^{129}$, which was abundant in meteorites. The $Xe^{129}$ itself resulted from the rapid neutron-capture process (the r-process) and was also thought to be produced in supernovae. Its abundance in meteorites thus implied important things about both the galactic environment and the formation of solar system. Cameron, strongly drawn to these issues, began a long and ardent quest to study and come to better understand them.

The launch of Sputnik (October 4, 1957) led to new U.S. enthusiasm for space research and to the creation of the National Aeronautics and Space Administration (NASA). By spring 1961, Cameron had become one of the first hires at NASA's Goddard Institute of Space Studies (GISS) in New York City. Directed by Robert Jastrow, GISS supported both junior and senior postdocs, and faculty on sabbatical, in wide-ranging areas of astrophysics and planetary science. It had a new supercomputer, an IBM 360-95, which at the time was faster than those at Los Alamos and Livermore and much less crowded with users. Though slower than today's cell phones (but considerably larger), it was still a dramatic change for academic scientists.

Cameron began a strong program of topical conferences at GISS, and he recruited graduate students from nearby universities (Columbia, NYU, and Yale at first), as described in (3). His conference organization extended to the early Texas conferences on relativistic astrophysics, the Gordon Research conferences on nuclear chemistry, and the introduction of astrophysics into the programs at the Aspen Center for Physics.

For an illustration of Cameron's research interests, consider his first generation of graduate students and their topics: Sachiko Tsuruta, neutron star structure and cooling; Robert Stein, solar hydrodynamics; Arnold Gilbert, reaction rates for complex nuclei; Carl Hansen, weak interactions in stars; James Truran, an approach to nuclear statistical equilibrium; Dave Arnett, core collapse and supernovae. Given that four of the six were directly related to issues of nucleosynthesis, supernova explosions and core collapse, and the formation of neutron stars and black holes, there were powerful synergies.

Cameron did not assign tasks; he offered options and gave the students freedom to learn independently. He believed that "once you have mastered the basics, you learn better when you find things out for yourself, and when you get to the point of teaching it you learn it still better." Application of this philosophy led to the Yale lecture notes (5), and to some extent the CRL-41 (Cameron 1957a). Cameron would first lecture his students on nuclear astrophysics to introduce them to their dissertation topics in a broad context, and then have them "explore the universe as seen through the eyes of a physicist" by reading his *Physics of the Solar System* and *Galactic* and *Stellar Physics*. Despite their imposing titles, these two basic titles were from the Course of Theoretical Physics given by L. D. Landau and E. M. Lifshitz, but with application to astronomy. It was all quite exhilarating.

While neutron stars had been predicted in 1939 by theorists J. Robert Oppenheimer and GeorgeVolkoff, they were not observed for decades (Jocelyn Bell discovered the first pulsar in 1967; the X-ray binary Cen-X $^3$ was detected by the Uhuru satellite in 1971). In one of his 1959 papers, Cameron had focused on the mass required to make a neutron star become a black hole. He found that nuclear forces would shift the maximum neutron star mass to a value above the Chandrasekhar limiting mass for white dwarfs, making neutron stars plausible as astronomical objects. How could they form? Much of the GISS work was related to this question.

Fred Hoyle had already identified the nucleus $Fe^{56}$ as being the ashes resulting from the last stage of nuclear burning, but the nature of iron peak and core collapse was poorly understood. Truran, Hansen, Cameron, and Gilbert (1966) performed the first serious calculation of this process (now called silicon burning), in which a network of reactions that follow the last stages of burning to iron was solved by computer. This work required estimates of the rates of all relevant reaction links, both forward and inverse, as well as development of new numerical methods to accurately solve the strongly coupled nonlinear equations. But if the iron would be swallowed by the collapse of the stellar core to form a neutron star or black hole, how were terrestrial and meteoritic iron formed? Assuming that such a core collapse would be accompanied by a supernova explosion, Truran, Arnett, and Cameron (1967) examined the synthesis of nuclei in the ejected matter, and they discovered that for such rapid burning $Ni^{56}$ was the dominant form of ashes. This nucleus decays to $Co^{56}$ and then to $Fe^{56}$, providing a source for terrestrial iron and (by radioactive heating) a source for supernova light curves. Bodansky, Clayton, and Fowler (1968) later found the same result using a different method (quasi-equilibrium), adding an alternate approach to the problem.

After Cameron taught a course in space physics at the Belfer Graduate School of Science at Yeshiva University in 1965, he was invited to join the faculty. GISS was becoming more rigid and bureaucratic, and the greater flexibility of academic life had its attractions (as well as the reduction of commuting, perhaps). The peak in government funding occurred at about this time, followed by more intense competition for grants. This did not directly impact Cameron, but it severely affected Belfer, whose future looked dim. In 1972, George Field asked Cameron if he was interested in a Harvard University appointment. At that time, Field was attempting to reorganize astronomy at Harvard, which consisted of the Harvard College Observatory and the Smithsonian Astrophysical Observatory. Among his innovations was an umbrella organization called the Center for Astrophysics (CFA), and he appointed Cameron as its associate director for planetary sciences.

"Space Science is Big Science, and it costs Big Public Money. As such it requires Big Justification, involving Big Planning and Big Advice," Cameron wrote in (4), and as that reference reveals, he was involved numerous times in such activities during his Harvard years, beginning with service on the 1970 Decadal Survey of Astronomy (Greenstein committee). Among other assignments, Cameron was also a member of the steering committee and chair of the Space Astronomy Panel, a member of the committee on planetary and lunar exploration (COMPLEX) of the National Academies' Space Science Board (SSB), and chair of the SSB. In 1983 he was awarded the NASA Distinguished Public Service Medal.

As the capabilities of computers grew dramatically during Cameron's career, so did his participation and virtuosity. When he arrived at the CFA its computer was a CDC 6400, an expensive and not especially versatile choice at that time. Cameron wanted computing to be powerful, interactive, and distributed, but he understood that the transition toward this goal would be an ongoing process. It began with minicomputers from the Digital Equipment Corporation and Data General, extended to microcomputers for word processing, then to Sun workstations, and then to Intel workstations running Windows NT. Cameron was an enthusiastic convert to each new stage, and loved to show off his latest new tool. He was a regular contributor to *BYTE* magazine.

The focus of Cameron's research shifted toward solar system and planetary science, and, because of his many service roles, toward collaboration. He worked on the structure of the solar nebula with Milton Pine, on stellar evolution with Dilhan Ezer, on the solar nebula equation of state with Richard Epstein and Jas Mercer-Smith, on giant planets with Morris Poldack, and on numerical hydrodynamics with Willy Benz. With the discovery of the "fractionated and unknown nuclear" (FUN) anomalies, Cameron became interested in the implied time constraints and triggered star formation. In order to provide radioactive species ($^{26}$Al, $^{36}$Cl, $^{41}$Ca, $^{53}$Mn, $^{60}$Fe) to the solar nebula, Truran and Cameron (1977) suggested that a supernova triggered the process.

The merger of Cameron's scientific and computing interests occurred in earnest in the mid-1970s, when he began to simulate the formation of the Earth-Moon system. The problem, being dynamic and three-dimensional, was beyond the capability of computers of the time, regarding both speed and cost. But before long he was happily simulating the effects of a giant impact with Earth, an activity that continued even into his retirement. He also learned how to do particle-based fluid dynamics by computer as he went, and by 1975 results were beginning to come in (Hartmann and Davies 1975; Cameron and Ward 1976). An important example of Cameron's work from this period is Benz, Slattery, and Cameron (1986), in which the researchers constrained the impactor mass and collision velocity. With increasing algorithmic sophistication and raw computer power, the understanding of the formation of the Earth-Moon system has since grown in complexity and depth (see Canup 2004, Stevenson 2014).

After retiring from Harvard in 1999, Alastair and his wife moved to Tucson, where he had been appointed senior research scientist at the Lunar and Planetary Laboratory of the University of Arizona. He and Betsy designed a unit in the Arizona Senior Academy, a retirement village conceived by the University's president Henry Koffler to attract retired academics who were still intellectually active. Located east of Tucson, with a view of the mountains but also a convenient shuttle to campus, the Camerons' unit was unique in its computer room, which was full of his favorite devices.

Over the course of his career, Cameron received many honors. They included the Canadian Astronomical Society's R. M. Petrie Prize Lectureship (1970), the NASA Distinguished Public Service Medal (1983), the J. Lawrence Smith Medal of the National Academy of Sciences (1988), the American Geophysical Union's Harry H. Hess Medal (1989), the Leonard Medal of the Meteoritical Society (1994), and the Henry Norris Russell Lectureship of the American Astronomical Society (1997).

Cameron died of a heart attack on October 3, 2005, on his way to a conference. It was just a few days after he learned he'd been awarded that year's Hans Bethe Prize of the American Physical Society.

Cameron had pursued his scientific research with great technical skill tempered by wisdom. We leave the reader with two quotes from (4) that suggest the man's style:

> *Physicists often exhibit a great deal of arrogance, since they consider that they are practicing the queen of the sciences. But I have learned that there is a great deal of value in what people in the other sciences have to tell me about their observations and conclusions. I reserve the option of reinterpreting what they have told me so that it fits together with whatever else I know about the subject. Sometimes this process will cause me to change my mind about something I thought I already understood.*

> *The core of my intellectual approach to trying to understand the universe is to seek consistency everywhere. Of course, ugly facts are always coming along to spoil beautiful theories, but sometimes the revised theories that incorporate the ugly facts are even more beautiful. Sometimes they make you realize that nature is complex and you do not really understand it. But I have a reputation for frequently changing my mind and that is caused by the constant search for consistency. I am counting on that to sustain me and keep me mentally alive as I head toward retirement. But although I will have a pension to replace my salary, I want to keep the computers running as I try to resolve yet another inconsistency, as long as I am physically able to do so.*

*This narrative was based on colleagues' personal recollections of Alastair Cameron, fortified and confirmed by the following sources:*

# SELECTED BIBLIOGRAPHY